\documentclass[%
twocolumn,
superscriptaddress,
showpacs,
bibnotes,
amsmath,amssymb,
aps,
prl,
]{revtex4-1}

\usepackage{graphicx} 
\usepackage{dcolumn} 
\usepackage{bm} 

\usepackage{ulem} 
\usepackage{color}

\usepackage{multirow}

\usepackage{array,ragged2e}
\usepackage{booktabs}

\begin{document}


\title{Plethora of Coexisting Topological Band Degeneracies in
  Nonsymmorphic Molecular Crystal OsOF$_5$}

\author{Hyungjun Lee}
\email{E-mail: hyungjun.lee@epfl.ch}
\affiliation{ %
  Institute of Physics, \'{E}cole
  Polytechnique F\'{e}d\'{e}rale de Lausanne (EPFL), CH-1015 Lausanne,
  Switzerland}
\affiliation{National Centre for Computational Design and Discovery of Novel Materials MARVEL, \'{E}cole Polytechnique F\'{e}d\'{e}rale de Lausanne (EPFL), CH-1015 Lausanne, Switzerland}
\author{Oleg V. Yazyev}
\email{E-mail: oleg.yazyev@epfl.ch}
\affiliation{ %
  Institute of Physics, \'{E}cole
  Polytechnique F\'{e}d\'{e}rale de Lausanne (EPFL), CH-1015 Lausanne,
  Switzerland}
\affiliation{National Centre for Computational Design and Discovery of Novel Materials MARVEL, \'{E}cole Polytechnique F\'{e}d\'{e}rale de Lausanne (EPFL), CH-1015 Lausanne, Switzerland}

\date{\today}

\begin{abstract}
In band theory of solids, degeneracies manifest themselves as point, line, and surface 
objects. The topological nature of band degeneracies has recently been appreciated since the 
recognition of gapless topological quantum phases of matter. 
Based on first-principles calculations, we report the topological semimetal phase in osmium oxyhalide OsOF$_5$ that exhibits zero-dimensional Weyl points, one-dimensional nodal lines, and two-dimensional nodal surfaces.
The semimetal phase is ensured by the electron filling constraint
while band degeneracies of various dimensionalities are protected by the
nonsymmorphic crystalline symmetries. Focusing on two nodal loops
crossing the Fermi level, we demonstrate their topological protection
by $\pi$ Berry phase and describe the drumhead surface states topologically connected by nontrivial $\mathbb{Z}_2$ index.
Our prediction highlights OsOF$_5$ as a unique topological material in which a rich variety of topological degeneracies exists in a narrow energy range close to the Fermi level owing to the molecular crystal structure of this material.
\end{abstract}


\maketitle

Symmetry is one of the fundamental guiding principles of
nature~\cite{Feynman1965}. In physics,
various kinds of discrete or continuous symmetries play a conscious
role as the guides to physical laws and phenomena, and serve as one of the organizing principles in their classification~\cite{Brading2003}.
Not only does its notion facilitate the more concise and elegant description of
physical systems and phenomena, but also its invariance and violation
give rise to the emergence of a host of physical
phenomena~\cite{Livio2012,Witten2018}. 
Symmetry constrains the allowable dynamics in physical
systems~\cite{Gross1996} and implies the conservation laws~\cite{Noether1918}.
Symmetry
breaking had been serving as an
organizing principle in the classification of possible phases of
matter and
transitions between them~\cite{Landau1980} before the recognition of topological order~\cite{Wen2017}.

Symmetry is also at the heart of the degeneracy in
electronic bands in crystalline solids.
Given a symmetry group under which the Hamiltonian of the system of
interest is invariant, we can use the symmetry to determine the
existence and order of degenerate levels~\cite{Dresselhaus2008}.
This study of degeneracies in crystals traces
back to the 
early days of the band theory of solids:
Bouckaert~\textit{et al.}~\cite{Bouckaert1936} tackled this problem
with an emphasis on the degeneracy necessitated by symmetry, the so-called
essential degeneracy, and Herring addressed another type of degeneracy
occurring in crystals, namely the accidental
degeneracy~\cite{Herring1937}, in a sense that degeneracy is not a
direct consequence of symmetry.

Recently, the subject of degeneracy in the energy bands of crystals has been reignited by the theoretical predictions of topological semimetal phases of
matter~\cite{Wan2011,Young2012}, in which the valence and conduction bands touch at discrete
points or continuous lines, forming low-dimensional Fermi surfaces. Similar to their insulating counterparts, topological (semi)metallic phases have topologically protected boundary modes, such as the surface Fermi arcs~\cite{Wan2011}, and display exotic
transport properties such as the chiral anomaly~\cite{Nielsen1983}.

Both gapped and gapless topological phases rely on the symmetry
of the system. For instance, the nontrivial topology in topological insulators (TIs)
is related to their time-reversal symmetry (TRS) $\mathcal{T}$ and that
of the
topological crystalline insulators pertains to the
crystalline symmetry. In the case of gapless topological
phases, Weyl semimetals (WSMs) require the translational symmetry and
Dirac semimetals (DSMs) need additional crystalline symmetries such as rotational symmetry to stabilize their band touching points~\cite{Armitage2018}.
In particular, crystalline
symmetries are diverse and one can expect a host of nontrivial topology
which originate from them~\cite{Shiozaki2014,Chiu2014}.
Among the symmetry-protected topological phases, topological nodal-line
semimetals, one of the possible topological semimetal
phases~\cite{Fang2016,Armitage2018}, attract much interest 
as a fertile ground for exploring the interplay between topology and
symmetry~\cite{ShuoYing2018}, as well as the intermediate
phase in the transitions to other topological phases such
as TIs~\cite{Weng2015prb}, WSMs~\cite{Huang2015,Weng2015} and DSMs~\cite{Kim2015,Yu2015}.

In this Letter, we report the discovery of a novel topological semimetal phase in
OsOF$_5$ 
enforced by its electron filling under the
combination of time-reversal and nonsymmorphic crystalline symmetries. The material
exhibits a rich variety of coexisting topological band degeneracies of
various dimensionalities close to the Fermi energy ($E_\mathrm{F}$), namely the
zero-dimensional (0D) Weyl points, one-dimensional (1D) nodal lines, and
two-dimensional (2D) nodal surfaces. 
Focusing on two nodal loops crossing $E_\mathrm{F}$, we demonstrate their nontrivial topology of $\pi$
Berry phase and describe the drumhead surface states topologically
connected by nontrivial $\mathbb{Z}_2$ index.
Our work adds to the family of topological materials a new member that is unique in a sense that the 
multitude of topological band degeneracies are confined 
within a narrow energy range near the Fermi level, a consequence of its molecular crystal 
structure, and hence novel exotic physical phenomena are expected to be discovered in OsOF$_5$.

\begin{figure}[t]
  \centering
  \includegraphics[width=8.5cm]{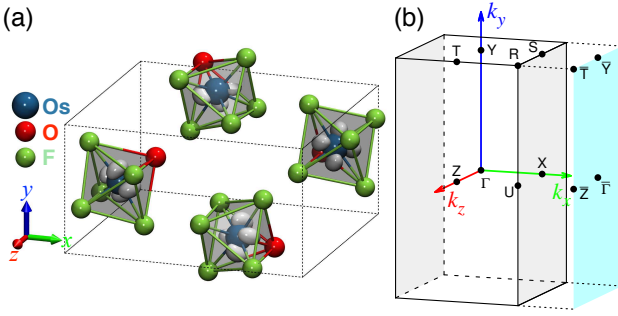}
  \caption{(a) Ball-and-stick model of the crystal structure of
    OsOF$_5$. 
Within each octahedron, the Wannier functions forming the isolated group of eight bands at the Fermi level are shown.
The primitive unit cell is outlined.
(b) Bulk Brillouin zone (BZ) and the cyan-shaded (100) surface BZ of
OsOF$_5$. High-symmetry points are indicated by black dots with the
corresponding labels.
}
  \label{fig:cryst}
\end{figure}

Figure~\ref{fig:cryst}(a) shows the crystal
structure of osmium oxide pentafluoride, OsOF$_5$. The material crystallizes in
the orthorhombic structure below 32.5$^\circ$ and there are two space
groups (SGs), SG \textit{Pnma}
(\#{}62)~\cite{Bartlett1968,Bartlett1968b} and SG \textit{Pna}2$_1$
(\#{}33)~\cite{Shorafa2006}, available in the literature.
Here, we focus on the more recent crystal
structure with SG \textit{Pna}2$_1$~~\footnote{We also investigated the
  structure with SG \textit{Pnma}, but the main conclusions of our work don't
  change. The differences are: (1) the nodal lines are fourfold
  degenerate and (2) there are no Weyl points and nodal surfaces. Both of
  them are attributable to the presence of inversion symmetry in SG \textit{Pnma}}.
OsOF$_5$ consists of four octahedral building
blocks with O and F atoms at
the vertices and Os in the center of each octahedron~\cite{Shorafa2006}. 
SG \textit{Pna}2$_1$ has no inversion symmetry $\mathcal{P}$ as its
element and is
generated by multiple nonsymmorphic symmetries, namely, two glide mirror planes, $\tilde{M}_x=\{M_x|(\frac{a}{2}\frac{b}{2}\frac{c}{2})\}$ and $\tilde{M}_y=\{M_y|(\frac{a}{2}\frac{b}{2}0)\}$,
and one screw rotation axis,
$\tilde{C}_{2z}=\{C_{2z}|(00\frac{c}{2})\}$, where $a$, $b$ and $c$
are the lattice constants along the $x$, $y$ and $z$ axes, respectively.
These
multiple nonsymmorphic
symmetries endow OsOF$_5$ with a rich variety of topological nodal structures~\cite{Bouhon2017}, as discussed in the rest of the paper.

\begin{figure}[t]
  \centering
  \includegraphics[width=7.5cm]{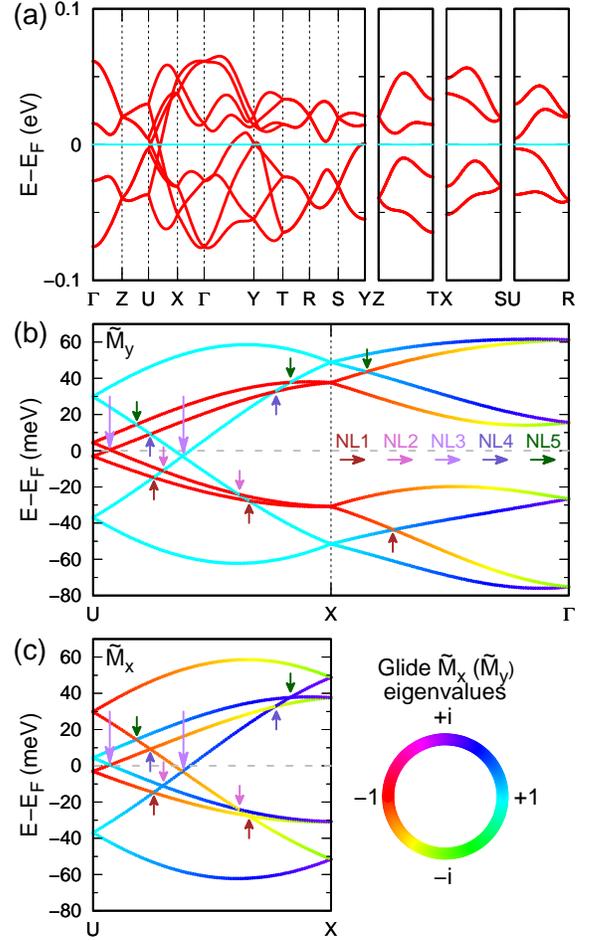}
  \caption{(a) Calculated bulk band structure of OsOF$_5$ showing the eight-band group at the Fermi level along the high-symmetry
    lines.
(b, c) Zoom-in plot of the band structure showing the details of band degeneracies due to the underlying
nonsymmorphic symmetries. All band touchings, part of the nodal-line
structures, are indicated by the arrows with the different colors representing the nearest band pairs. Additionally, band structures
are color-coded according to the eigenvalues of the glide symmetry
    operators, (b) $\tilde{M}_y$ and (c) $\tilde{M}_x$.
}
  \label{fig:band}
\end{figure}

Figure~\ref{fig:band}(a) shows the bulk band structure of OsOF$_5$
near $E_\mathrm{F}$ along several high-symmetry lines obtained from first-principles calculations with spin-orbit
coupling (SOC) included.
Details of the computational methodology are given in the Supplemental Material~\footnote{See Supplemental Material at http://link.aps.org/supplemental/... for the details of the computational methods.}.
Following the experimental
results~\cite{Bartlett1968}, we study the paramagnetic phase of OsOF$_5$.
As can be seen in Fig.~\ref{fig:band}(a), an isolated group of eight
bands spans the energy region around $E_\mathrm{F}$. This eight-band
manifold with mainly Os $d_{xy}$-orbital character, as seen in Fig.~\ref{fig:cryst}(a), is the lowest of five isolated eight-band groups, all of which have Os $d$
character. The energy gap between the discussed eight-band group
crossing $E_\mathrm{F}$ and the second-lowest group is about
0.46~eV. The set of five groups is separated from the continuum bands
with mainly O or F character by 2.6~eV above and 1.1~eV below.
Hereafter, we limit the discussion only to the eight-band manifold crossing
$E_\mathrm{F}$ with the bandwidth of 0.14~eV.

There are two important features of the band structure presented in
Fig.~\ref{fig:band}(a). The first one is the semimetal phase of
OsOF$_5$. It is a molecular
crystal, and thus its band structure can be expected to consist of fully populated and empty
isolated groups of bands exhibiting a gapped phase. Contrary to this naive expectation,
OsOF$_5$ features a semimetal phase ensured by electron filling in conjunction with
its nonsymmorphic SG~\cite{Watanabe2016}. 
Such restrictive filling constraints were recently applied for
the screening of topological semimetal candidates from a large database of materials~\cite{Chen2018}.
From the electron-filling criterion it follows that OsOF$_5$ hosts
a gapless phase at filling $\nu=8n+4$.

The second remarkable feature is the connected eight-band
manifold. A group of bands is called ``connected'' if one can travel
continuously through all its branches owing to the touching points
between them~\cite{Michel1999}.
To clearly demonstrate it, we plot the band dispersions along the
U-X-$\Gamma$ and the U-X lines [Figs.~\ref{fig:band}(b) and (c),
respectively]. 
The presence of the connected eight-band manifold and its robustness can be explained by symmetry arguments. 
All \textit{k} points along the U-X line are invariant under three nonsymmorphic symmetry operations  -- $\tilde{M}_x$, $\tilde{M}_y$, and
$\tilde{C}_{2z}$. Thus, the
energy eigenstates can be labeled simultaneously with the eigenvalues
of these nonsymmorphic symmetry operators.
On the other hand, for \textit{k} points along the $\Gamma$-X line,
only $\tilde{M}_y$ belongs to the little co-group. 
On the glide-invariant planes, the eigenvalues of $\tilde{M}_x$ and $\tilde{M}_y$, denoted as $\tilde{m}_x$ and $\tilde{m}_y$, can be written as
$\tilde{m}_x(\mathbf{k})=\pm ie^{-i(k_yb+k_zc)/2}$ and $\tilde{m}_y(\mathbf{k})
=\pm ie^{-ik_xa/2}$\,. It follows that for the time-reversal-invariant momentum (TRIM)
points $\Gamma$ and X on
the $\Gamma$-X line, the twofold-degenerate Kramers partners have the opposite eigenvalues $\tilde{m}_y=\pm\,i$ at $\Gamma$, but at $X$ they
should have the same eigenvalue $\tilde{m}_y=+1$ or
$\tilde{m}_y=-1$ owing to the Kramers degeneracies between the
complex-conjugate representations at the TRIM points. This results in partner switching between the two TRIM points due to the continuity of
glide eigenvalues, and thus leads to an
unavoidable band crossing 
between $\Gamma$ and X, forming the connected ``four-band'' group with the
hourglass-shaped dispersion, the typical building-block of the band structure of nonsymmorphic crystals with SOC and TRS~\cite{Young2015,Wieder2016b,Bzdusek2016}.

In order to confirm the fact that instead of four bands, eight bands should
be tangled together in SG \textit{Pna}2$_1$, one has to check additionally the evolution of glide
eigenvalues along the X-U line where $\tilde{M}_x$ and $\tilde{M}_y$
are the elements of the little group.
While $\tilde{m}_y$ is constant with the value
of $+1$ or $-1$, $\tilde{m}_x$ along the X-U line shows the same hourglass-type evolution. Importantly, due to the continuity of the glide
eigenvalues, the partners of the states should be exchanged again
along this line, resulting
in a band dispersion that represents two overlapping
hourglasses~\cite{Wieder2016b}.
The reasoning above is also confirmed by the evolution of glide
eigenvalues shown in Figs.~\ref{fig:band}(b) and (c), which are directly calculated from \textit{ab initio} Bloch states.
It is this type of eight-band connectivity that explains why $\nu=8n+4$ implies a semimetal
phase~\cite{Watanabe2016} for SG \textit{Pna}2$_1$.
Bearing in mind this along with the fact that the
level repulsion can be avoided between the states belonging to different
irreducible representations, we can confirm that there are
irremovable band touching points along these high-symmetry lines arising due to the
nonsymmorphic symmetries.

\begin{figure}[t]
  \centering
  \includegraphics[width=7.5cm]{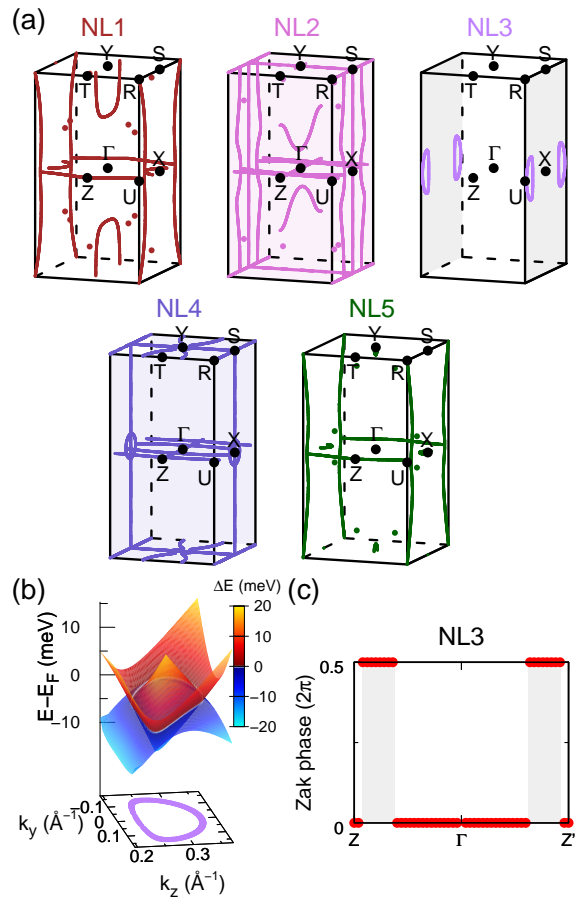}
  \caption{(a)  Band degeneracies of different dimensionalities in OsOF$_5$ formed by band
    crossings between bands $(N-2)$ and $(N-1)$ (NL1), 
    $(N-1)$ and $N$ (NL2), 
    $N$ and $(N+1)$ (NL3), $(N+1)$ and $(N+2)$ (NL4), and, finally, bands $(N+2)$ and $(N+3)$ (NL5). 
    Here, $N$ is the number of valence electrons in the unit cell. In NL2 (NL4), the nodal
    surfaces at $k_z=\pm\pi/c$ are highlighted by shaded planes. (b) Band dispersion around
    the nodal loop (NL3). Bands are
    encoded by the color corresponding to the energy difference
    between the lowest conduction and highest valence bands. The
    contour of the nodal loop is plotted underneath. (c) Evolution of Zak phase for the nodal loop (NL3) along the high-symmetry line from
    Z $(0,0,\pi/c)$ to Z' $(0,0,-\pi/c)$.}
  \label{fig:fig3}
\end{figure}

We can expect that these band degeneracies are part of
symmetry-protected nodal-line structures on the glide-invariant planes~\cite{Bzdusek2016} including the $\Gamma$-X or X-U line.
In order to verify this suggestion and identify the complete
structure of degeneracies, we performed the direct search of
all possible band crossings. 
We found band degeneracies of different dimensionalities such
as 0D Weyl points, 1D nodal lines, and finally 2D nodal surfaces, as
summarized in Fig.~\ref{fig:fig3}(a).  
Their detailed descriptions are in order.
First, there are two nodal loops on the BZ boundary with
$k_x=\pm\pi/a$,
which result from the twofold degeneracy between bands $N$ and $(N+1)$
(NL3), where $N$ is the number of valence electrons in the unit cell.
These nodal loops cross $E_\mathrm{F}$ with almost no dispersion in energy.
Second, between bands $(N-2)$ and $(N-1)$ (NL1), as well as between
bands $(N+2)$ and $(N+3)$ (NL5), there are nodal chains in addition to
nodal loops on the $k_x=0$ and $k_y=0$ planes.
In NL1 and NL5 we also found 12 and 8 Weyl points, respectively, all of which have the
Chern number of $\pm 1$. Considering the present symmetries, the number of
symmetry-inequivalent Weyl points are 3 and 2, respectively.
Third, between bands $(N-1)$ and $N$ (NL2), and bands $(N+1)$ and
$(N+2)$ (NL4), we found more complex band degeneracies
composed of extended nodal lines, nodal loops and finally essential nodal
lines and surfaces.
We conclude that the nodal surfaces with $k_z=\pm\pi/c$ are protected by
the composite symmetry $\mathcal{T}\cdot\tilde{C}_{2z}$ with SOC and no inversion symmetry~\cite{Liang2016,Wu2018}.

The robustness of Weyl points can be proved by their non-zero
chirality and that of nodal surfaces is guaranteed by their origin
from essential degeneracy. Thus, we only need to confirm the
stability of nodal lines. 
Their robustness is ensured by nonsymmorphic symmetries of OsOF$_5$ as
described above
and they can be further topologically protected by the $\pi$ Berry
phase~\cite{Chan2016}. For this purpose, we calculated the Zak phase
to integrate the Berry potential along a reciprocal vector
$\mathbf{b}_1$ (parallel to $k_x$) starting
from the \textit{k} points on the $k_x=0$ plane. 
As can be seen in
Fig.~\ref{fig:fig3}(c), the Zak phase for the NL3 changes
discontinuously from 0 to $\pi$ across the nodal line, yielding the $\pi$ Berry phase
for any rectangular loop interlinking with it. This
demonstrates the topological protection of nodal rings (NL3) crossing $E_\mathrm{F}$.

\begin{figure}[t]
  \centering
  \includegraphics[width=7cm]{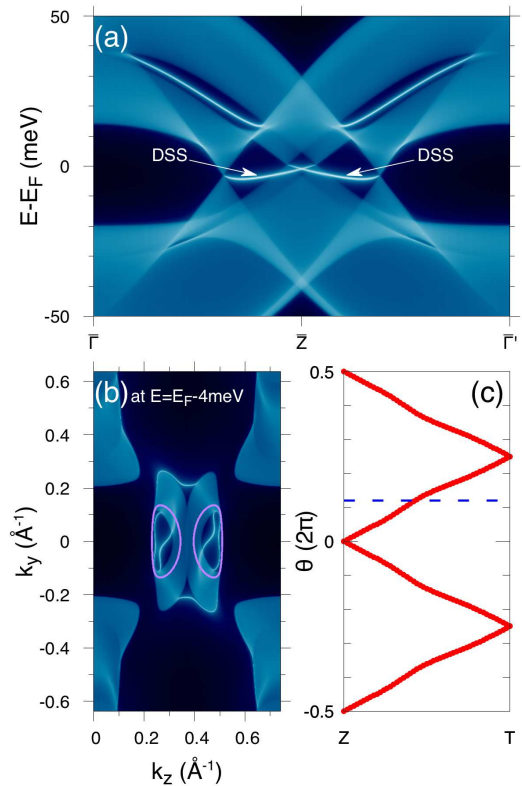}
  \caption{(a) Momentum-resolved local density of states at the
    (100) surface of OsOF$_5$. The drumhead surface states (DSSs)
    are indicated by white arrows. (b) Constant-energy
    map of (100) surface bands at $E=E_\mathrm{F}-4\,\mathrm{meV}$.
    The projections of the nodal loops (NL3) onto the (100) surface BZ are
    indicated by purple lines. (c) Evolution of Wilson-loop
    eigenvalues $\theta$ for the half-filled eight-band manifold on
    the gapped TRS-invariant plane with $k_z=\pi/c$. The odd number of crossing
  with the reference line indicates the nontrivial $\mathbb{Z}_2$
  index.}
  \label{fig:surfacestates}
\end{figure}

The quantized $\pi$ Berry phase is also closely related to the surface
states of topological nodal-line semimetals~\cite{Burkov2011,Chan2016}. It is known that
in contrast to other topological phases, the boundary states of topological nodal-line
semimetals are not fully protected in that the surface states can be
pushed out of the gap and merged into the bulk states~\cite{Fang2016,Burkov2018},
but their connectivity with the projections of the
nodal lines is protected~\cite{Burkov2018}, forming the so-called
drumhead surface states~\cite{Burkov2011,Kopnin2011}. 
Their presence characteristic of topological nodal-line semimetals can
lead to a topological polarization~\cite{Ramamurthy2017} and a possible surface superconducting order originating from a large density of states due to their weak energy dispersion~\cite{Kopnin2011}.

Figure~\ref{fig:surfacestates}(a) shows the calculated surface-state
band dispersion revealing nearly flat drumhead surface states,
indicated by the arrows, which cross $E_\mathrm{F}$ and connect the projections of the nodal lines. To reveal
more clearly the drumhead surface states, we calculated the
constant-energy map of surface bands at 4~meV below
$E_\mathrm{F}$. The surface states are visible inside the projections
of the nodal loops [Fig.~\ref{fig:surfacestates}(b)].
In general, the drumhead surface states are expected to be confined
within the projection of the nodal lines onto the surface BZ, but their band dispersion and connectivity are very sensitive
to the details of the surface structure~\cite{Bian2016prb,Bian2016}. 
As seen in Fig.~\ref{fig:surfacestates}(a),
two surface states are linked across the zone border and this
behavior can also be seen in Fig.~\ref{fig:surfacestates}(b). 
Two drumhead surface states are connected through the bulk states and
this connectivity is topologically protected by nontrivial
$\mathbb{Z}_2$ index~\cite{Kane2005Z2} on the gapped TRS-invariant planes~\cite{Bzdusek2016} between two nodal loops, as seen in~Fig.~\ref{fig:surfacestates}(c).

To conclude, we report a complex topological semimetal phase in OsOF$_5$ and investigate it in detail by performing first-principles calculations. A series of band degeneracies of different dimensionalities--Weyl points, nodal lines and nodal surfaces--are shown to coexist in a partially filled eight-band manifold.
The unique feature of this material is that these topological band degeneracies are confined within a narrow energy range close to the Fermi level, a consequence of its molecular crystal structure.
While being a new member of the emerging family of topological semimetals, OsOF$_5$
can also be related to the class of organic charge-transfer molecular crystals in which Mott insulating, superconducting, spin liquid and other phases \cite{Powell2011} along with the topologically protected band degeneracies \cite{Kobayashi2007,Commeau2017} have been extensively investigated. Similarly, we expect novel exotic physical phenomena to be observed in OsOF$_5$.

\begin{acknowledgments}
  We thank D. Gos\'{a}lbez-Mart\'{i}nez for helpful discussions.
  This work was supported by the
  European Research Council starting grant ``TopoMat'' (Grant
  No. 306504) and the NCCR MARVEL, funded by the Swiss National Science Foundation. First-principles computations have been performed at
  the Swiss National Supercomputing Centre under Project No. s832.
\end{acknowledgments}

%

\end{document}